\documentclass[
 aip,
amsmath,amssymb,
 reprint,%
]{revtex4-1}

\usepackage{graphicx}
\usepackage{dcolumn}
\usepackage{bm}

\usepackage[utf8]{inputenc}
\usepackage[T1]{fontenc}
\usepackage{mathptmx}
\usepackage{etoolbox}

\makeatletter
\def\@email#1#2{%
 \endgroup
 \patchcmd{\titleblock@produce}
  {\frontmatter@RRAPformat}
  {\frontmatter@RRAPformat{\produce@RRAP{*#1\href{mailto:#2}{#2}}}\frontmatter@RRAPformat}
  {}{}
}%
\makeatother
\begin{document}

\preprint{}

\title[A highly-sensitive broadband superconducting thermoelectric single-photon detector]{A highly-sensitive broadband superconducting thermoelectric single-photon detector}

\author{Federico Paolucci}
\email[Author to whom correspondence should be addressed: ]{federico.paolucci@pi.infn.it}
\affiliation{ 
INFN Sezione di Pisa, Largo B. Pontecorvo 3, I-56127, Pisa, Italy}

\author{Gaia Germanese}
\affiliation{ 
NEST, Istituto Nanoscienze-CNR and Scuola Normale Superiore, I-56127 Pisa, Italy}
\affiliation{ 
Dipartimento di Fisica dell’Universit\`a di Pisa, Largo B. Pontecorvo 3, I-56127, Pisa, Italy}

\author{Alessandro Braggio}
\affiliation{ 
NEST, Istituto Nanoscienze-CNR and Scuola Normale Superiore, I-56127 Pisa, Italy}

\author{Francesco Giazotto}
\affiliation{ 
NEST, Istituto Nanoscienze-CNR and Scuola Normale Superiore, I-56127 Pisa, Italy}
\email{francesco.giazotto@sns.it}

\date{\today}

\begin{abstract}
We propose a passive single-photon detector based on the bipolar thermoelectric effect occurring in tunnel junctions between two different superconductors thanks to spontaneous electron-hole symmetry breaking. Our thermoelectric detector ($TED$) converts a finite temperature difference caused by the absorption of a single photon into an open circuit thermovoltage.
Designed with feasible parameters, our $TED$ is able to reveal single-photons of frequency ranging from $\sim$15 GHz to $\sim$150 PHz depending on the chosen design and materials. In particular, this detector is expected to show values of signal-to-noise ratio $SNR\sim15$ at $\nu=50$ GHz when operated at a temperature of 10 mK. Interestingly, this device can be viewed as a \emph{digital} single-photon detector, since it generates an almost constant voltage $V_S$ for the full operation energies.
Our $TED$ can reveal single photons in a frequency range wider than 4 decades with the possibility to discern the energy of the incident photon by measuring the time persistence of the generated thermovoltage.
Its broadband operation suggests that our $TED$ could find practical applications in several fields of quantum science and technology, such as quantum computing, telecommunications, optoelectronics, THz spectroscopy and astro-particle physics. 
\end{abstract}

\maketitle
The detection of single quanta of electromagnetic radiation, i.e. photons, plays a fundamental role in countless applications in modern quantum science and technology. To reveal single photons of frequency lower than 1 THz, cryogenic detector are necessary, due to the low energy released during their absorption. 
In particular, the temperature dependent electronic properties of superconducting devices are exploited to reveal low energy single photons with large efficiency and high energy resolution \cite{Semenov2002}. Indeed, transition-edge sensors ($TES$s) \cite{Irwin1995}, kinetic inductance detectors ($KID$s) \cite{Day2003} and superconducting nanowire single-photon detectors ($SNSPD$s) \cite{Gol2001} represent the state-of-the-art for practical technological applications \cite{Kraus1996}. To further push the sensitivity towards lower frequencies, devices exploiting the Josephson effect \cite{Josephson1962} have been envisioned in the form of superconducting tunnel junctions ($STJ$s) \cite{Oelsner2013,Guarcello2019}, proximitized nanosystems (metallic or semiconducting) \cite{Giazotto2008,Govenius2016,Lee2020}, Josephson escape sensors ($JES$s)\cite{Paolucci2020}, temperature-to-phase conversion ($TPC$) \cite{Virtanen2018} and temperature-to-voltage conversion ($TVC$) \cite{Solinas2018} detectors. These systems take advantage of miniaturization to increase the thermal decoupling between the charge carriers and the thermal bath, thus reducing their thermal capacitance and boosting their detection sensitivity. On the contrary, the readout operation of these devices implemented by an external bias may perturb the frail superconducting state with detrimental impact on the detection performance. Furthermore, powering extended arrays of cryogenic detectors might be even problematic, since each bias line carries a large amount of heat into the system limiting the total number of operative devices.  As a consequence, passive superconducting detectors generating electrical signals after the photon absorption are expected to boost the sensitivity towards lower frequencies. To this scope, sensors based on the thermoelectricity predicted to occur in superconducting/ferromagnetic tunnel junctions by explicitly breaking the particle-hole (PH) symmetry \cite{Ozaeta2014} have been proposed \cite{Hei2018,Giazotto2015}. Despite the great interest, the experimental realization of such devices will be limited by the difficulties in the synthesis of suitable materials. 

\begin{figure}
\centering
\includegraphics[width=1\columnwidth]{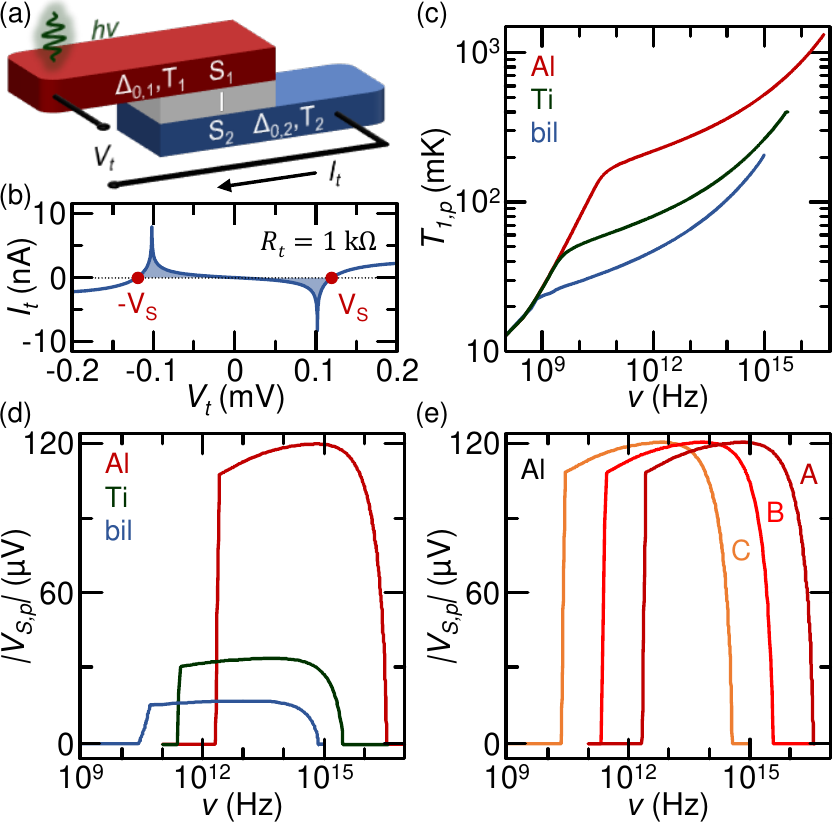}
\caption{\label{Fig1} (a) Schematic representation of the superconducting thermoelectric detector $TED$ consisting of a S$_1$IS$_2$ tunnel junction (with $\Delta_{0,1}>\Delta_{0,2}$). The absorption of a photon of frequency $\nu$ increases the electronic temperature of S$_1$ ($T_1>T_2$). The voltage drop ($V_t$) and the circulating current ($I_t$) are shown. (b) Thermoelectric $I_t$ vs. $V_t$ characteristic calculated for $\Delta_{0,1}=210\;\mu$eV, $r=0.5$, $\Gamma_{i}=10^{-4}\times\Delta_{0,i}$ (with $i=1,2$) and $R_t=1$ k$\Omega$ at $T_1=840$ mK and $T_2=10$ mK. The thermovoltages $\pm V_S$ are represented. The blue shaded areas represent the thermoactive regions. (c) Maximum electronic temperature $T_{1,p}$ reached by S$_1$ versus $\nu$ calculated for a volume $\mathcal{V}_1=0.25\;\mu$m$^{-3}$ made of Al (red), Ti (green) and an Al/Cu bilayer (blue) at $T_b=10$ mK. (d) Thermovoltage $|V_{S,p}|$ generated at the photon absorption versus $\nu$ considering the same system of panel (c), $r=0.5$ and $T_2=10$ mK. (e) $|V_{S,p}|$ versus $\nu$ calculated for an Al $TED$ with S$_1$ of volume $\mathcal{V}_{1,A}=0.25\;\mu$m$^{-3}$, $\mathcal{V}_{1,B}=2.5\times10^{-2}\;\mu$m$^{-3}$ and $\mathcal{V}_{1,C}=2.5\times10^{-3}\;\mu$m$^{-3}$.}
\end{figure}

Here, we propose a passive single-photon termoelectric detector ($TED$) that does not require any external power source \cite{Germanese2023}. The detector works as follows: a lead absorbs the electromagnetic energy $h\nu$ (with $h$ the Planck constant and $\nu$ the frequency) of a single photon and increases its electronic temperature. Consequently, an open circuit voltage is spontaneously generated by the system. To this end, our device takes advantage of the bipolar thermoelectric effect theoretically predicted \cite{Marchegiani2020,Marchegiani2020b,Marchegiani2020c} and experimentally demonstrated \cite{Germanese2022,Germanese2022b} in tunnel junctions between two different superconductors subject to large temperature gradients (non-linear response regime) if the Josephson coupling is sufficiently suppressed \cite{Marchegiani2020d}. This $TED$ can operate in a wide range of photon frequency (from 10 GHz to 10 PHz) thus finding applications in several fields of quantum technology.

The core of our $TED$ is a S$_1$IS$_2$ tunnel junction with completely suppressed Josephson coupling (via Fraunhofer effect with an in-plane magnetic field or by implementing the junctions in a Josephson interferometer), where S$_1$ and S$_2$ are the two superconductors (with the zero-temperature energy gaps obeying to $\Delta_{0,1}>\Delta_{0,2}$) and I is the insulating barrier [see Fig. \ref{Fig1}(a)]. In this system, the charge transport is due to quasiparticles and is described by \cite{Tinkham}
\begin{equation}
I_{t}=\frac{1}{eR_t}\int_{-\infty}^{+\infty}\text{d}\epsilon N_{1}(\epsilon )N_{2}(\epsilon+eV_{t})[f_1(\epsilon)-f_2(\epsilon+eV_t)],
\label{eq:iqp}
\end{equation}
where $R_t$ is the junction normal-state resistance, $\epsilon$ is the energy and  $f_{1,2}(\epsilon)=1/[\exp(\epsilon/k_BT_{1,2})+1]$ (with $k_B$ the Boltzmann constant and $T_{1,2}$ the electron temperature) is the Fermi distribution in the $1,2$ lead. For Bardeen-Cooper-Schrieffer (BCS) superconductors, the density of states takes the form $N_{1,2}(\epsilon)=|\text{Re}[(\epsilon+i\Gamma_{1,2})/\sqrt{(\epsilon+i\Gamma_{1,2})^2-\Delta_{1,2}^2}]|$, where $\Delta_{1,2}$ is the temperature-dependent superconducting energy gap and $\Gamma_{1,2}$ is the Dynes broadening parameter \cite{Dynes1984} of the $1,2$ lead. Within these assumptions, the S$_1$IS$_2$ junction generates thermoelectric signals for $T_1/T_2 \gtrsim \Delta_{0,1}/\Delta_{0,2}$ (with $T_1$ and $T_2$ the electronic temperature of S$_1$ and S$_2$ , respectively) and $r=\Delta_{0,1}/\Delta_{0,2}\lesssim 0.83$ \cite{Marchegiani2020,Germanese2022}. In the following, we set a value of the gaps ratio $r=0.5$ providing optimal thermoelectric performance. Indeed, the current versus voltage characteristic in Fig. \ref{Fig1}(b) shows the typical absolute negative conductance ($I_tV_t<0$) of thermoelectric generation calculated from Eq. \ref{eq:iqp} for $\Delta_{0,1}=210\;\mu$eV, $\Gamma_{i}=10^{-4}\times\Delta_{0,i}$ (with $i=1,2$), $R_t=1$ k$\Omega$, $T_1=840$ mK and $T_2=10$ mK (under thermal gradient). 
Due to the intrinsic PH symmetry of the two superconducting leads, this system presents a reciprocal $I_tV_t$ characteristics [$I_t(V_t)=-I_t(-V_t)$] and two possible stable values of thermovoltage $\pm V_S$ [such that $I_{t}(\pm V_s)=0$] for a given temperature gradient thanks to non-equilibrium spontaneous particle-hole symmetry breaking \cite{Marchegiani2020,Germanese2022,strocchi}. For our analysis, we will only focus on the positive branch of thermoelectricty, but our device spontaneously selects a positive or negative solution. In the following, we will assume to not distinguish between the two cases by considering the absolute value of the generated signal.

To exploit the S$_1$IS$_2$ system as a $TED$, we need to evaluate the maximum value ($T_{1,p}$) reached by $T_1$ due to the absorption of a single photon of frequency $\nu$. To achieve this goal we need to solve \cite{Mosley1984,Chui1992}
\begin{equation}
    \int_{T_1=T_b}^{T_{1,p}(\nu)} C_{e}(T) \text{d}T = h\nu,
\end{equation}
where $T_b$ is the phonon bath temperature and $C_e(T)= T \tfrac{\text{d}}{\text{d}T} \mathcal{S}_1 (T)$ is the electronic thermal capacity of S$_1$. The bulk superconductor entropy of S$_1$ reads \cite{Rabani2008}
\begin{equation}
\mathcal{S}_1 (T)=-2\mathcal{V}_1 \mathcal{N}_F k_B \int_{-\infty}^{\infty} \text{d}\epsilon N_{1}(\epsilon,T)f_1(\epsilon,T)\ln[f_1(\epsilon,T)],
\end{equation}
where $\mathcal{V}_1$ is the volume and $\mathcal{N}_F$ is the density of states at the Fermi energy. Figure \ref{Fig1}(c) shows $T_{1,p}$ versus $\nu$ calculated at $T_b=10$ mK for S$_1$ of volume $\mathcal{V}_1=0.25\;\mu$m$^{-3}$ made of aluminum ($\Delta_{0,Al}=210\;\mu$eV, red), titanium ($\Delta_{0,Ti}=60\;\mu$eV, green) and an aluminum/copper bilayer ($\Delta_{0,bil}=30\;\mu$eV, blue). As expected, the peak temperature of S$_1$ monotonically increases with the frequency, and large superconducting gap materials ensure higher values of $T_{1,p}$ at a given $\nu$ stemming from a reduced heat capacitance. Then, this temperature increase can be employed to calculate the open circuit thermovoltage $V_{S,p}$ originated by the photon absorption. To this end, we solve Eq. \ref{eq:iqp} for $T_{1,p}(\nu)$ and $T_2=T_b=10$ mK (S$_{2}$ is supposed to be fully thermalized with the bath), as shown in Fig. \ref{Fig1}(d). 

On the one hand, the generated $V_{S,p}$ lowers by decreasing the energy gap of the superconductor forming S$_1$\cite{Marchegiani2020}, since $V_S\sim[\Delta_{1}(T_1)-\Delta_{2}(T_2=T_b)]/e$. 
On the other hand, small gap superconductors ensure sensitivity to photons of lower frequency having smaller electronic thermal capacitance. Indeed, the thermoelectric generation requires a finite minimum and maximum temperature gradients both scaling with the energy gap of S$_1$ \cite{Marchegiani2020}. 
Notably, the $TED$ efficiently reveals photons over four frequency decades irrespective of the material composing S$_1$, thereby our structure realises a broadband single-photon $TED$ with an \emph{almost constant} output voltage.
As a consequence, this device can be seen as a \emph{digital} broadband superconducting thermoelectric single-photon detector. In this perspective, our $TED$ strongly differs from the other superconducting thermoelectric detectors \cite{Hei2018,Giazotto2015}, where $V_S$ depends linearly on the temperature reached by the absorber, i.e. the energy of the incoming photon.
Futhermore, for a given material, the operation window of the detector can be shifted towards lower frequencies (but with constant $V_{S,p}$) by decreasing the volume of S$_1$ (and thus its thermal capacitance $C_e$), as shown in Fig. \ref{Fig1}(e) for aluminum.

\begin{figure}
\centering
\includegraphics[width=1\columnwidth]{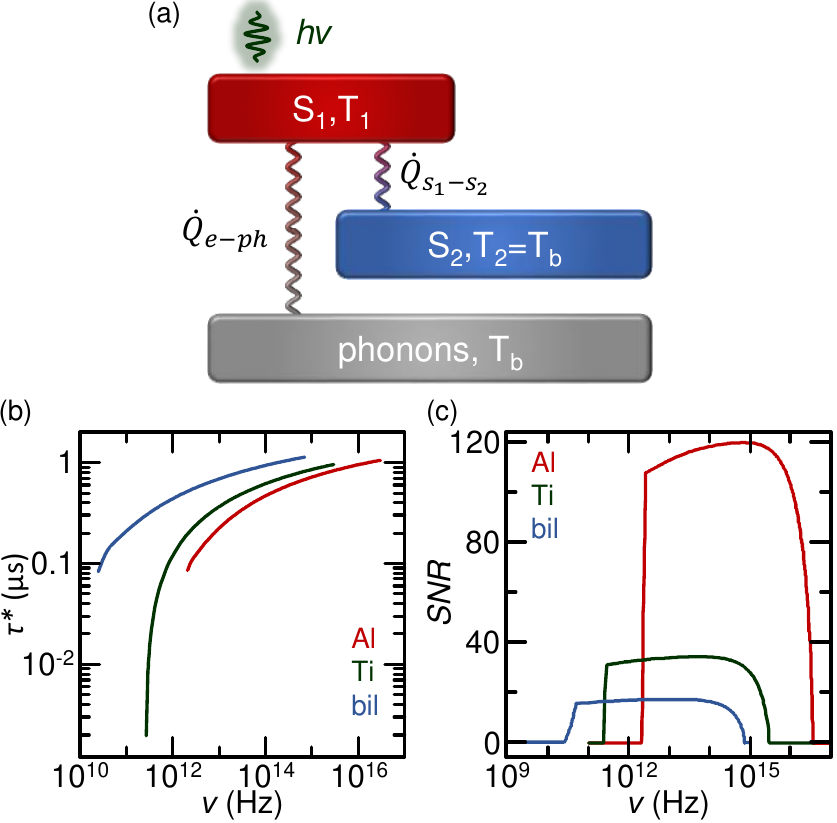}
\caption{\label{Fig2} (a) Thermal model of the $TED$: the energy $h\nu$ of the incoming photon causes the rise of $T_1$. The active region S$_1$ thermalises back to $T_b$ thanks to electron-phonon heat exchange $\Dot{Q}_{e-ph}$ and quasiparticle heat current through the junction $\Dot{Q}_{S_1-S2}$. Detector time constant $\tau^*$ (b) and signal-to-noise-ratio $SNR$ (c) versus photon frequency ($\nu$) calculated for $S_1$ of volume $\mathcal{V}_1=0.25\;\mu$m$^{-3}$ made of Al (red), Ti (green) and an Al/Cu bilayer (blue) at $R_t=100\;\Omega$, and $T_b=10$ mK.}
\end{figure}

To theoretically address the performance of our single photon $TED$, we need evaluate the predominant electronic heat transport channels of S$_1$, as shown in Fig. \ref{Fig2}(a). After photon absorption, the quasiparticle thermalization in S$_1$ is described by $\Dot{Q}_{tot}(T_1,T_b)=\Dot{Q}_{e-ph}(T_1,T_b)+\Dot{Q}_{S_1-S_2}(T_1,T_b)$, where $\Dot{Q}_{e-ph}(T_1,T_b)$ accounts for the electron-phonon heat exchange and $\Dot{Q}_{S_1-S_2}(T_1,T_b)$ represents the thermal current flowing through the junction [see Fig. \ref{Fig2}(a)]. In a superconductor, the electron-phonon thermalization is exponentially damped by the presence of the energy gap, and it can be approximated by \cite{Timofeev2009,Heikkila2019}
\begin{equation}
\begin{aligned} 
    \Dot{Q}_{e-ph}(T_1,T_b)= \delta T G_{ph}(T_1)
    \approx \delta T \frac{\Sigma_1 \mathcal{V}_1 T_1^4}{96 \zeta(5)}\times \\
    \Big[ \mathfrak{f}_1 \left(\frac{k_B T_1}{\Delta_1}\right)e^{-\Delta_1/k_B T_1}+
    \pi \left(\frac{\Delta_1}{k_B T_1}\right)^5 \mathfrak{f}_2 \left( \frac{k_B T_1}{\Delta_1}\right)e^{-2\Delta_1/k_B T_1} \Big],
    \end{aligned}
    \label{eq:el-ph}
\end{equation}
where $\delta T=T_1-T_b$ is the temperature difference between electrons and phonons, $\Sigma_1$ is the electron-phonon coupling constant, $\zeta(5)$ is the Riemann zeta function, $\mathfrak{f}_1(x)=\sum_{n=0}^3 C_n x^n$ (with $C_0 \approx 440$, $C_1 \approx 500$, $C_2 \approx 1400$, $C_3 \approx 4700$) and $\mathfrak{f}_2(x)= \sum_{n=0}^2 C_n x^n$ (with $B_0=$64, $B_1=$144, $B_2=$258). Yet, the heat transport through the S$_1$IS$_2$ junction takes the form \cite{Maki1965,Golubev2013}
\begin{equation}
\begin{aligned} 
\Dot{Q}_{S_1-S_2}(T_1,T_b)=\frac{1}{e^2 R_t}\times\\
\int_{-\infty}^{+\infty}\text{d}\epsilon \;\epsilon_S N_{1}(\epsilon,T_1)N_{2}(\epsilon_S,T_b)[f_1(\epsilon,T_1)-f_2(\epsilon_S,T_b)],
\end{aligned} 
\label{eq:ithqp}
\end{equation}
where $\epsilon_S=\epsilon+eV_s$. The timescale of the thermalization process in S$_1$ can be estimated by the equation $C_e(T_1) \frac{dT_1}{dt}=-\Dot{Q}_{tot}(T_1,T_b)$. Consequently, the $TED$ characteristic time $\tau^*$ representing the persistence of signal generation after the single photon absorption reads
\begin{equation}
    \tau^*(\nu,T_b)=\int_{T_1^*}^{T_{1,p}(\nu)} \text{d}T_1 \frac{C_e(T_1)}{\Dot{Q}_{tot}(T_1,T_b)},
    \label{eq:taustar}
\end{equation}
where $T_1^*$ is the minimum temperature enabling thermoelectricity ($V_{S}\neq0$).

Figure \ref{Fig2}(b) shows the dependence of $\tau^*$ on $\nu$ at $T_b=10$ mK and $R_t=100\;\Omega$ for different materials. In particular, $\tau^*$ ranges between 1 ns to 1 $\mu$s for all photon frequencies and chosen materials. At a given photon frequency, the $TED$ produces an output signal for a longer time for small gap superconductors.
On the one hand, larger values of $\tau^*$ increases the \emph{blind} time between detection of two photons. On the other hand, persistent output signals demand a slower electronics with advantages in terms of noise bandwidth and cost. We note that, despite $V_s$ has a relatively weak dependence on $\nu$ [see Fig. \ref{Fig1}(d)], the recovery time $\tau^*$ (thermovoltage generation time) allows us to discern the energy of the incoming photon over a wide range of frequency. Thus, our $TED$ can operate as a broadband single-photon spectrometer.

The detector signal-to-noise-ratio $SNR$ provides the strength of the output signal relative to the noise. In our $TED$ it is defined as
\begin{equation}
SNR[T_{1,p}(\nu)]=\frac{V_{S,p}[T_{1,p}(\nu)]}{S_{V,amp}\sqrt{\omega}},
\label{eq:SNR}
\end{equation}
where $S_{V,amp}=1$ nV$/\sqrt{\text{Hz}}$ is the voltage noise spectral density of a typical low-noise voltage pre-amplifier used to read the open circuit thermovoltage, and $\omega=1$ MHz is the bandwidth. We assume no contribution by the junction itself, since in the absence of incident radiation the detector doesn't generate \emph{any} signal. 
Figure \ref{Fig2}(c) shows the $\nu$-dependence of the $SNR$ calculated for different materials at $T_b=10$ mK. In the operation frequency window, the $TED$ shows remarkable values of the signal-to-noise ratio reaching up to about 120 for aluminum. At a given radiation frequency and bandwidth, the $SNR$ scales with the superconducting energy gap of S$_1$, since it is proportional to $V_{S,p}[T_{1,p}(\nu)]$ (see Eq. \ref{eq:SNR}). 

\begin{figure}
\centering
\includegraphics[width=1\columnwidth]{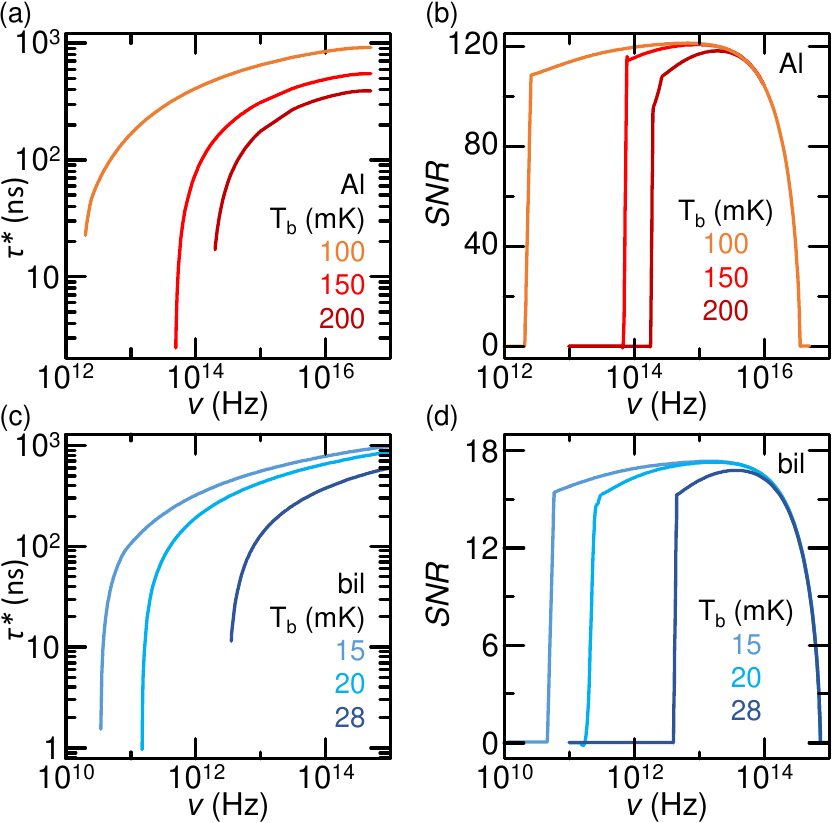}
\caption{\label{Fig3} $TED$ time constant $\tau^*$ (a) and signal-to-noise-ratio $SNR$ (b) versus $\nu$ calculated for S$_1$ made of Al at different values of the bath temperature $T_b$. $\tau^*$ (c) and $SNR$ (d) versus $\nu$ calculated for S$_1$ made of Al/Cu bilayer at different values of the bath temperature $T_b$. All the curves are calculated for $r=0.5$, $\Gamma_{i}=10^{-4}\times\Delta_{0,i}$ (with $i=1,2$), $\mathcal{V}_1=0.25\;\mu$m$^{-3}$, and $R_t=100\;\Omega$.}
\end{figure}

We now focus on the dependence of the $TED$ performance in bath temperature assuming that the quasiparticles in S$_2$ fully thermalize with the phonons bath ($T_2=T_b$). To this end, we consider the same device structure presented in the base temperature behavior, namely $r=0.5$, $\Gamma_{i}=10^{-4}\times\Delta_{0,i}$ (with $i=1,2$), $\mathcal{V}_1=0.25\;\mu$m$^{-3}$ and $R_t=100\;\Omega$. Here, the S$_1$ superconductor is realized in the form of aluminum or Al/Cu bilayer, as shown in Fig. \ref{Fig3}, thus allowing us to explore the complete frequency range of detectable single photons by our $TED$. On the one hand, the operation speed of the device increases by rising the bath temperature, as shown by the drop of $\tau^*$ with $T_b$ by both aluminum [panel (a)] and Al/Cu bilayer [panel (c)]. 
The latter is due to the increase of the quasiparticle thermalization $\Dot{Q}_{tot}$ with bath temperature (see Eqs. \ref{eq:el-ph}, \ref{eq:ithqp} and \ref{eq:taustar}). We stress that the $\tau^*$ versus $\nu$ characteristic is still monotonic, thus ensuring the spectrometer operation of our single-photon $TED$. On the other hand, both the best $SNR$ and the frequency sensitivity range of our $TED$ reduce by increasing $T_b$, as shown by panels (b) and (d) of Fig. \ref{Fig3} for aluminum and Al/Cu bilayer, respectively. At a given $T_b/T_{C1}$ ratio (with $T_{C1}$ the critical temperature of S$_1$), the operation single photon frequency window decreases more for low gap superconductors, because thermoelectricity is weaker \cite{Marchegiani2020}.

In summary, we have proposed an original highly-sensitive broadband passive digital superconducting single-photon detector. To this scope, we exploit the recently discovered bipolar thermoelectricity in superconducting S$_1$IS$_2$ tunnel junctions due to spontaneous particle-hole symmetry breaking. Possible practical implementations of the $TED$ suppressing the detrimental Josephson coupling \cite{Marchegiani2020,Marchegiani2020b,Marchegiani2020c} include superconducting quantum interference devices \cite{Germanese2022,Germanese2022b} and extended tunnel junctions immersed in an in-plane magnetic field\cite{Rowell1963,Perez2014}. 
Our detector is able to reveal single-photons of frequency ranging from $\sim$15 GHz to $\sim$150 PHz depending on the choice of the superconducting material. 
For each material, the device produces an almost constant $V_{S,p}$ irrespective of the frequency of the incident radiation, thus it can be viewed as a \emph{digital} superconducting thermoelectric single-photon detector.
Our $TED$ shows values of signal-to-noise ratio $SNR\sim15$ at $\nu=50$ GHz for a sensitive element made of a superconductor of zero-temperature energy gap $\Delta_0=30\;\mu$eV. Interestingly, this detector can reveal single photons over a broadband frequency window wider than 4 orders of magnitude. Furthermore, our $TED$ can work as spectrometer by measuring the time persistence $\tau^*$ of the generated thermovoltage after the photon absorption.

\begin{figure}
\centering
\includegraphics[width=1\columnwidth]{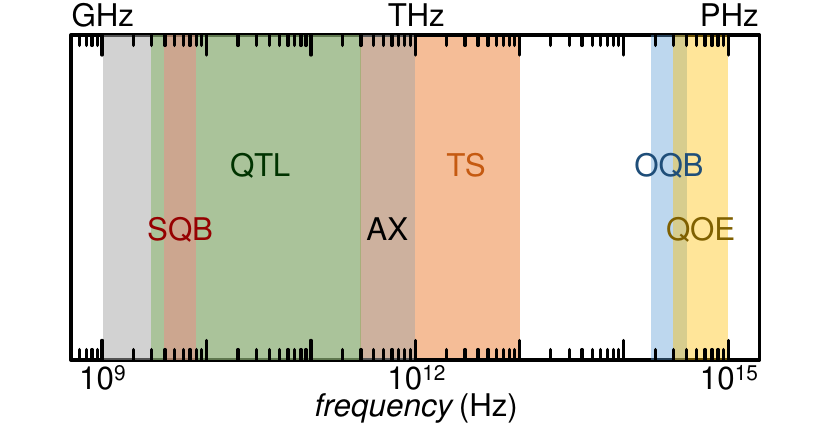}
\caption{\label{Fig4} Frequency range of possible applications of our single photon $TED$: superconducting qubits (SQBs), quantum telecommunications (QTLs) axions search (AX), THz specroscopy (TS), optical qubits (OQBs) and quantum opto-electronics (QOE).}
\end{figure}

This broadband $TED$ could find practical exploitation in several fields of quantum science and technology where the management of single photons is fundamental, as schematically shown in Fig. \ref{Fig4}. In particular, possible applications encompass fundamental science, such as the search of axions (AXs)\cite{Sikivie1983,Capparelli2016} in laboratory experiments and THz spectroscopy (TS) \cite{Peiponen} of new materials and molecules, and quantum technologies, such as telecommunications (QTLs) \cite{Chen2021}, optoelectronics (QOE)\cite{Ossiander2022}, and superconducting (SQBs)\cite{Siddiqi2021} or optical (OQBs)\cite{Slussarenko2019} qubits. 

Finally, the composition of our $TES$ can be also generalized by exploiting different materials and structures. Indeed, a tunnel junction between a hot electrode characterized by a gapped density of states and a cold lead with a monotonically decreasing density of states is expected to show thermoelectricity due to PH symmetry breaking\cite{Marchegiani2020}. For instance, bipolar thermoelectricty was recently predicted to occur in tunnel junctions between a bilayer graphene and a superconductor \cite{Bernazzani2022}. Furthermore, the exploitation of semiconductors and two-dimensional crystals might enable the operation of this family of detectors towards higher temperatures, thus improving their versatility and widening their application area.  

\begin{acknowledgments}
The authors wish to thank G. Marchegiani and C. Guarcello for fruitful discussion. We acknowledge the EU’s Horizon 2020 research and innovation program under Grant Agreement No. 800923 (SUPERTED), No. 964398 (SUPERGATE) and No. 101057977 (SPECTRUM) for partial financial support. A.B. acknowledges the Royal Society through the International
Exchanges between the UK and Italy (Grants No. IEC
R2 192166 and IEC R2 212041)
\end{acknowledgments}

\section*{Data Availability Statement}
The data that support the findings of this study are available from the corresponding author upon reasonable request.

\bibliography{aipsamp}

\end{document}